\documentclass[twoside]{article}
\usepackage{fleqn,espcrc2}
\usepackage{graphicx}
\usepackage{amssymb}

\newcommand{\AmS}{{\protect\the\textfont2
  A\kern-.1667em\lower.5ex\hbox{M}\kern-.125emS}}

\title{Statistical Analysis of Solar Neutrino Data\thanks{
Talk presented by C. Giunti at
NOW 2000, Conca Specchiulla (Otranto, Italy), 9-16 Sep. 2000;
DFTT 46/00, hep-ph/0012247.}
}

\author{M.V. Garzelli and C. Giunti\\[0.2cm]
INFN, Sez. di Torino, and Dip. di Fisica Teorica,
Univ. di Torino, I--10125 Torino, Italy}
       
\begin{document}

\begin{abstract}
We calculate with Monte Carlo the goodness of fit
and the confidence level of the standard allowed regions
for the neutrino oscillation parameters
obtained from the fit of the total rates
measured in solar neutrino experiments.
We show that they are significantly overestimated
in the standard method.
We also calculate exact allowed regions
with correct frequentist coverage.
We show that
the exact VO, LMA and LOW regions are much larger than the standard ones
and merge together
giving an allowed band at large mixing angles
for all $\Delta{m}^2 \gtrsim 10^{-10} \, \mathrm{eV}^2$.
\end{abstract}

\maketitle

\section{Introduction}
\label{Introduction}

The data of solar neutrino experiments
provide strong indications in favor of neutrino oscillations
and their statistical analysis
give indications on the values of the neutrino mixing parameters.
For two neutrino generations,
these parameters are
the mass-squared difference
$\Delta{m}^2 \equiv m_2^2 - m_1^2$,
where $m_1$ and $m_2$ are the two neutrino masses,
and
$\tan^2\theta$,
where
$\theta$
is the mixing angle
(see, for example, Ref.\cite{BGG-review-98-brief}).

In this paper we present statistical methods
based on Monte Carlo numerical calculations
that allow
to improve the standard statistical analysis
of solar neutrino data,
which is approximate
\cite{Garzelli-Giunti-sf-00}.
We consider the data relative to the total rates
measured in the Homestake
and Super-Kamiokande
experiments,
and the weighted average of the total rates measured
in the two Gallium experiments
GALLEX and SAGE,
as given in Table I of Ref.\cite{Concha-sun-99-brief}.

As in the standard method,
we use as estimator of
$\Delta{m}^2$
and
$\tan^2\theta$
the global minimum
$X^2_{\mathrm{min}}$
of the least-squares function
\begin{eqnarray}
X^2
&=&
\sum_{j_1,j_2=1}^{N_{\mathrm{exp}}}
\left( R^{\mathrm{(thr)}}_{j_1} - R^{\mathrm{(exp)}}_{j_1} \right)
(V^{-1})_{j_1j_2}
\nonumber
\\
&&
\hspace{1cm}
\times
\left( R^{\mathrm{(thr)}}_{j_2} - R^{\mathrm{(exp)}}_{j_2} \right)
\,,
\label{X2}
\end{eqnarray}
where
$N_{\mathrm{exp}}=3$
is the number of experimental data points,
$V$ is the covariance matrix of
experimental and theoretical uncertainties,
$R^{\mathrm{(exp)}}_{j}$
is the event rate measured in the $j^{\mathrm{th}}$ experiment
and
$R^{\mathrm{(thr)}}_{j}$
is the corresponding theoretical event rate,
that depends on
$\Delta{m}^2$ and $\tan^2\theta$.
We calculate the covariance matrix $V$
following the standard method
(see, for example, Ref.\cite{Concha-sun-99-brief}),
with the correction proposed in
Ref.\cite{Garzelli-Giunti-cs-00}.

The standard procedure to calculate
the allowed regions in the neutrino oscillation parameter space
(see, for example, Ref.\cite{Concha-sun-99-brief})
is based on the assumption that
$X^2$
has a $\chi^2$ distribution with $N_{\mathrm{exp}}$ degrees of freedom.
This would be correct if
the theoretical rates
$R^{\mathrm{(thr)}}_{j}$
depended \emph{linearly} on the parameters
to be determined in the fit
and the errors of
$R^{\mathrm{(thr)}}_{j}-R^{\mathrm{(exp)}}_{j}$
were \emph{multinormally} distributed
with \emph{constant} covariance matrix $V$.
In this case there should be only one minimum of $X^2$
and the allowed regions at
$100\beta\%$ confidence level (CL),
given by
$X^2 \leq X^2_{\mathrm{min}} + \Delta{X^2}(\beta)$,
should be elliptical.

It is well-known
that in reality
there are several local minima of $X^2$,
each one determining an allowed region,
and
these allowed regions do not have elliptic form
(see, for example, Ref.\cite{Concha-sun-99-brief}).
This is due to the fact that the requirements above
are not satisfied.
In particular,
the stronger effect is due to
the non-linear dependence of the theoretical rates
$R^{\mathrm{(thr)}}_{j}$
from the parameters,
which generates
several local minima of $X^2$.

In the following two sections we present estimations
of the goodness of fit
(Section~\ref{GOF})
and
the confidence level of the standard allowed regions
(Section~\ref{CL})
obtained with a Monte Carlo
calculation of the distribution of $X^2$.
In Section~\ref{Exact}
we present the results of a Monte Carlo calculation
of allowed regions with exact coverage.

\begin{figure*}[tb]
\begin{center}
\begin{tabular*}{\textwidth}{@{\extracolsep{\fill}}lr}
\rotatebox{90}{\includegraphics[bb=120 80 555 741,width=6cm]{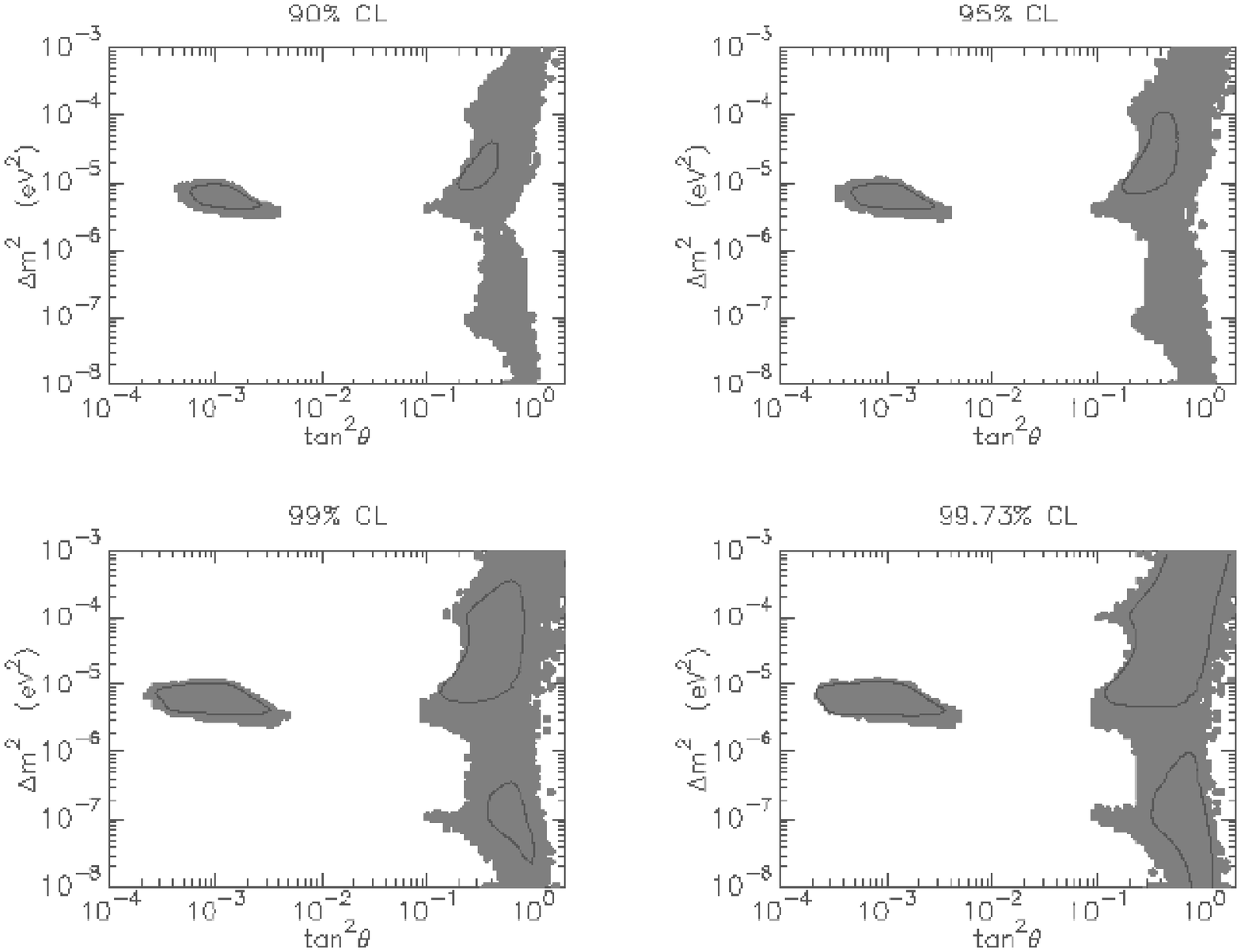}}
&
\rotatebox{90}{\includegraphics[bb=120 309 555 741,width=6cm]{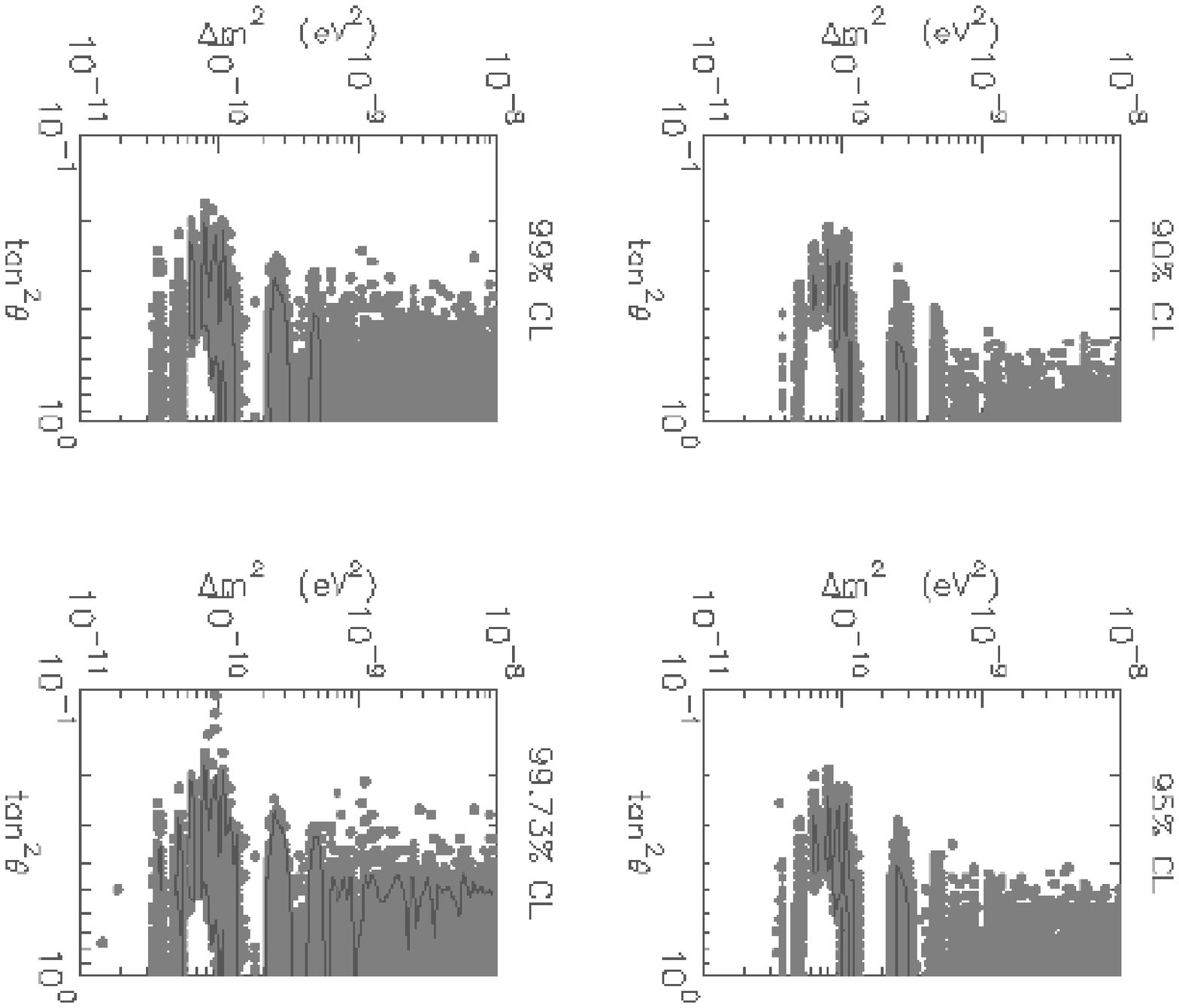}}
\end{tabular*}
\end{center}
\caption{ \label{cfi}
Allowed 90\%, 95\%, 99\%, 99.73\% CL regions.
The gray areas are the allowed regions with exact frequentist coverage.
The areas enclosed by the solid lines
are the standard allowed regions.
}
\end{figure*}

\section{Goodness of Fit}
\label{GOF}

Goodness of fit (GOF)
is the probability to find a value of the global minimum
$X^2_{\mathrm{min}}$
of $X^2$
larger than the one obtained from the fit.
Since there are more than one
local minima of $X^2$ with relatively close values of $X^2$,
there are more possibilities to obtain
good fits of the data
with respect to the case of one minimum,
and the true goodness of fit
is smaller than the one obtained with the standard method.

We calculate the distribution of
$X^2_{\mathrm{min}}$
assuming that the best-fit values
$\widehat{\Delta{m}^2}$, $\widehat{\tan^2\theta}$
of
$\Delta{m}^2$, $\tan^2\theta$
are reasonable surrogates of the true unknown values
$\Delta{m}^2_{\mathrm{true}}$, $\tan^2\theta_{\mathrm{true}}$
and
the probability distribution of the differences
$\widehat{\Delta{m}^2}_{(k)} - \widehat{\Delta{m}^2}$,
$\widehat{\tan^2\theta}_{(k)} - \widehat{\tan^2\theta}$
is not too different from the true distribution of the differences
$\widehat{\Delta{m}^2}_{(k)} - \Delta{m}^2_{\mathrm{true}}$,
$\widehat{\tan^2\theta}_{(k)} - \tan^2\theta_{\mathrm{true}}$
in a large set of best-fit parameters
$\widehat{\Delta{m}^2}_{(k)}$, $\widehat{\tan^2\theta}_{(k)}$
($k=1,2,\ldots$)
obtained with hypothetical experiments.

Using $\widehat{\Delta{m}^2}$, $\widehat{\tan^2\theta}$
as surrogates of the true values,
we generate $N_s$ synthetic random data sets
with the usual gaussian distribution for the
experimental and theoretical uncertainties.
We apply the least-squares method
to each synthetic data set,
leading to an ensemble of simulated best-fit parameters
$\widehat{\Delta{m}^2}_{(s)}$, $\widehat{\tan^2\theta}_{(s)}$
with $s=1,\ldots,N_s$,
each one with his associated
$(X^2_{\mathrm{min}})_{s}$.
Then we calculate the goodness of the fit
as the fraction of simulated
$(X^2_{\mathrm{min}})_{s}$
in the ensemble that are larger than the one actually observed,
$X^2_{\mathrm{min}}$.

The global minimum of the least-squares function (\ref{X2}),
$X^2_{\mathrm{min}} = 0.42$,
occurs in the SMA region\footnote{
We use the standard terminology for the allowed regions
(see, for example, Ref.\cite{Concha-sun-99-brief}):
SMA for
$\Delta{m}^2 \sim 5 \times 10^{-6} \, \mathrm{eV}^2$,
$\tan^2 \theta \sim 10^{-3}$,
LMA for
$\Delta{m}^2 \sim 3 \times 10^{-5} \, \mathrm{eV}^2$,
$\tan^2 \theta \sim 0.3$,
LOW for
$\Delta{m}^2 \sim 10^{-7} \, \mathrm{eV}^2$,
$\tan^2 \theta \sim 0.5$,
VO for
$\Delta{m}^2 \lesssim 10^{-8} \, \mathrm{eV}^2$.
}
for
$\Delta{m}^2 = 5.1 \times 10^{-6} \, \mathrm{eV}^2$
and
$\tan^2 \theta = 1.6 \times 10^{-3}$.
The Monte Carlo method yields a GOF of 40\%,
that must be compared with the 52\% GOF obtained with the standard method.
Therefore,
the standard method significantly overestimates the GOF.

\section{Confidence Level of Allowed Regions}
\label{CL}

The allowed regions with
$100\beta\%$ CL
are defined by the property that they belong to a
set of allowed regions,
obtained with hypothetical experiments,
which cover (\textit{i.e.} include) the true value of the parameters
with probability $\beta$.
This property is called \emph{coverage}.

When there are several local minima of $X^2$
with relatively close values of $X^2$,
in repeated experiments
the global minimum has significant chances
to occur far from the true (unknown)
value of the parameters,
leading to a smaller probability that the allowed regions
cover the true value
with respect to the case in which there is only one minimum,
which is assumed for the validity of
the standard method for the calculation of allowed regions.
Hence,
the true confidence level of a standard
$100\beta\%$ CL allowed region is smaller than $\beta$.

We estimate the true confidence level
of the standard $100\beta\%$ CL allowed regions
using the best fit values of the parameters,
$\widehat{\Delta{m}^2}$, $\widehat{\tan^2\theta}$,
as surrogates of the true values,
$\Delta{m}^2_{\mathrm{true}}$, $\tan^2\theta_{\mathrm{true}}$,
for the generation of a large number of synthetic data sets.
We apply the standard procedure to each synthetic data set
and obtain the corresponding standard $100\beta\%$ CL allowed regions
in the space of the neutrino oscillation parameters.
Then we count the number of synthetic standard
$100\beta\%$ CL allowed regions
that cover the assumed surrogate
$\widehat{\Delta{m}^2}$, $\widehat{\tan^2\theta}$
of the true values.
The ratio of this number and the total number of synthetically
generated data set gives a Monte Carlo estimation
$\beta_{\mathrm{MC}}$
of the true confidence level
of the standard $100\beta\%$ CL allowed regions.

We obtained that the confidence level of the standard
90\% CL allowed regions is 86\%,
which is significantly smaller.
Other results are presented in Ref.\cite{Garzelli-Giunti-sf-00}.

\section{Exact Allowed Regions}
\label{Exact}

The calculation
of the confidence level of the standard allowed regions
presented in the previous section
is approximate,
because it is based on the assumption of a surrogate
for the unknown true values of the neutrino oscillation parameters.
It would be useful to be able to calculate exact allowed
with the desired confidence level,
\textit{i.e.} with correct coverage.
The procedure that allows to
perform this task has been invented by Neyman in 1937
(see references in Ref.\cite{Garzelli-Giunti-sf-00}).
Let us emphasize that this procedure
gives allowed regions with proper coverage
for any unknown true values of the parameters.

Our implementation of Neyman's procedure
for the calculation of the allowed regions
in the
$\tan^2\theta$--$\Delta{m}^2$
plane
is described in details in Ref.\cite{Garzelli-Giunti-sf-00}.
The results are presented in Fig.\ref{cfi},
where the gray areas are the exact allowed regions,
and the standard allowed regions are enclosed by solid lines.
One can see that the exact
LMA, LOW and VO regions
are much larger than the standard ones
and
there is no separation between them.
Hence,
large mixing angles with
$0.2 \lesssim \tan^2\theta \lesssim 1$
are allowed for $\Delta m^2 \gtrsim 10^{-10} \, \mathrm{eV}^2$.
On the other hand,
the exact SMA region approximately coincides with the standard one.
This is due to the fact that in the SMA region
the assumption of a linear dependence of
the theoretical rates
$R^{\mathrm{(thr)}}_{j}$
in Eq.(\ref{X2})
from the parameters
$\tan^2\theta$, $\Delta m^2$
is approximately correct,
whereas it is violated quite badly
in the LMA, LOW and VO regions.

\section{Conclusions}
\label{Conclusions}

In conclusion,
we have shown that the standard method
used in the analysis of solar neutrino data
in terms of neutrino oscillations
significantly overestimates the goodness of fit
and the confidence level of the allowed regions.
We have also calculated allowed regions
with correct coverage.
The SMA region approximately coincides with the standard one,
but
the VO, LMA and LOW regions are much larger than the standard ones
and merge together
giving an allowed band at large mixing angles
for all $\Delta{m}^2 \gtrsim 10^{-10} \, \mathrm{eV}^2$.


\end{document}